\documentclass[showpacs,twocolumn,aps,prl]{revtex4-1}
\usepackage{graphicx,amsmath,amsfonts,amssymb}
\usepackage{subfigure}
\usepackage{bm}

\usepackage[
colorlinks=true,
linkcolor=blue,
urlcolor=blue,
citecolor=blue,
]{hyperref}

\begin{document}
\title{Strongly enhanced nonlinear acoustic valley Hall effect in tilted Dirac materials}

\author{Jia-Liang Wan$^{1}$}
\author{Ying-Li Wu$^{1}$}
\author{Ke-Qiu Chen$^{1}$}
\author{Xiao-Qin Yu$^{1}$}
\email{yuxiaoqin@hnu.edu.cn}
\affiliation{$^{1}$ School of Physics and Electronics, Hunan University, Changsha 410082, China}


\begin{abstract}
It has been recently established that a nonlinear valley current could be generated through traveling a surface acoustic wave (SAW) in two-dimensional Dirac materials. So far, the SAW-driven valley currents have been attributed to  warping Fermi surface or Berry phase effect. Here, we demonstrate that tilt mechanism can also lead to a nonlinear valley Hall current (VHC) when propagating SAW in materials with the tilted Dirac cone placed on a piezoelectric substrate. It's found that the nonlinear VHC exhibits a $\sin\theta$ dependence on the orientation of tilt with respect to SAW.
 In addition, 
this tilt-induced nonlinear acoustic VHC shows independence on the relaxation time, distinguishing from the contributions from the Berry phase or trigonal warping. Remarkably, the magnitude of the nonlinear acoustic VHC from tilt mechanism in the uniaxially strained graphene is two orders larger than those reported in MoS$_2$ stemmed from the Berry phase effect and the warping effect.
\end{abstract}

\maketitle
\label{Introduction}
\emph{Introduction}. The valley, an extra degree of freedom of electron, in two-dimensional (2D) crystal with honeycomb lattice structure shows potential to store and carry information instead of electron and spin, leading to the emergence of valleytronics \cite{Xiao2012,K.F.Mak2014,Shimazaki2015,Schaibley2016,Wu2019}, in which the generation of a valley current is a vital issue. The major approaches, nowadays, to generate a valley current are through the valley Hall (Nernst) effect \cite{Xiao2012,Yu2015}, which indicates electrons with different valleys ($\mathrm{K}$ and $\mathrm{-K}$ valleys) flowing in the opposite direction perpendicular to an applied electric field (temperature gradient) without breaking time-reversal symmetry. The generated valley current shows a linear dependence on the driven forces and can be attributed to a nonvanishing Berry curvature  \cite{Xiao2010Rev,Nagaosa2010Rev} of all occupied energy bands.

Recently, nonlinear anomalous Hall effect \cite{Sodemann2015,Low2015,Du2018,Du2019,Battilomo2019,Facio2018,Du2021} in time-reversal invariant noncentrosymmetric materials as a second-order response to an electric field, which stems from the dipole moment of Berry curvature near the Fermi level (namely Berry curvature dipole) \cite{Sodemann2015,Low2015,Du2018,Du2019,Battilomo2019,Facio2018,Du2021}, has attracted broad interests in studying on other  nonlinear anomalous transport phenomena, such as nonlinear spin Hall effect \cite{Hamamoto2017,araki2018}, nonlinear thermal Hall effect \cite{Zhou2022,Zeng2020RC} and the nonlinear anomalous Nernst effect \cite{Yu2019RC,Zeng2019,Wu2021}. All those effects are related to a  geometric property of electron wavefunctions, namely Berry curvature near the Fermi level, and driven by an electric field or temperature gradient.

\begin{figure*}[ht]
	\centering
		\includegraphics[width=1.0\linewidth]{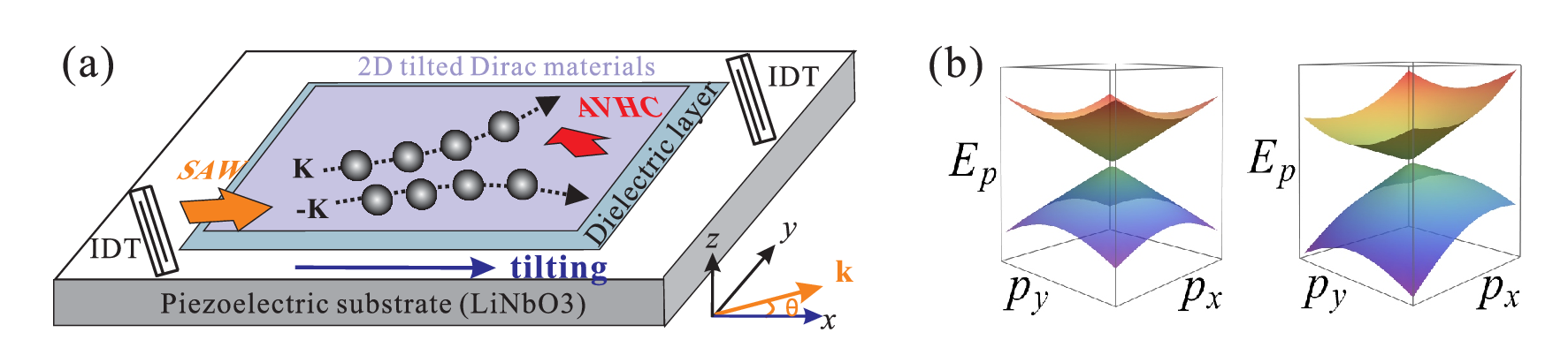}
	\caption{(a) Illustration of generation of VAHE through SAW for tilted  mechanism in a 2D material placed on a piezoelectric substrate. $\theta$ is the azimuthal angle of surface acoustic wave vector $\mathbf{k}$ with respect to the tilting direction (${x}$ direction).
(b) The band structure of the two-dimensional Dirac materials with [right] or without [left] tilting in $p_{x}$ direction.}
\label{Fig1}
\end{figure*}

In addition to the electron flows driven by an electric field and temperature gradient, acoustic waves can, actually, also drive carriers and generate an electric current through interaction with electrons. The acoustoelectric effect (AEE) \cite{Parm1953,Weinreich1957}, referring to a generation of electric current in response to the traveling acoustic wave, was firstly theoretically proposed by Parmenter \cite{Parm1953} in 1953 and observed in experiment by Weinreich \textit{et al.} in 1957 \cite{Weinreich1957}. The standard AEE originates in the sound-induced strain field and corresponding deformation potential which perturbs and drags electrons resulting in an electric current along the acoustic wave vector. Apart from the deformation potential mechanism, a piezoelectric mechanism of interaction between surface acoustic waves (SAW) and electrons has also been explored in low-dimensional systems (LDS) \cite{Wixforth1989,Willett1990,Fal1993,Hern2018,PerDelsing2019Rev}. When placing the LDS on the piezoelectric substrate, the Bleustein-Gulyaev (BG) acoustic wave generated through the interdigital transducers (IDTs) [Figure~\ref{Fig1}(a)] will induce a piezoelectric field and distort the ionic lattice, resulting in a local imbalance of electric chemical potential $\mu$ and leading to density fluctuation and nonequilibrium electron distribution. Consequently, the induced piezoelectric field drags carriers and gives rise to electron current.

Owing to the appearance of new two dimensional (2D) materials, the studies of SAWs are stimulated. The interactions with electrons have been investigated in monolayer graphene \cite{Zhang2011,Miseikis2012}, the surface of the topological insulators \cite{Parente2013}. And a few new acoustoelectric response have been recently predicted, including acoustic drag effect \cite{Kovalev2015}, valley acoustoelectric effect in two-dimensional transition metal dichalcogenides (TMDs), and pseudoelectromagnetic-field induced acoustogalvanic effect in Dirac and Weyl materials\cite{Sukhachov2020,Bhalla2022,Zhao2022}. Among them, a nonlinear acoustoelectric valley Hall effect (AVHE) as second-order response to the SAW-induced field stemming from piezoelectric or deformation potential mechanisms was reported in TMDs placed on a piezoelectric substrate \cite{Kalameitsev2019} or nonpiezoelectric substrate \cite{Sonowal2020}, respectively. For the deformation potential mechanism\cite{Sonowal2020}, the warping Fermi surface is crucial to get a nonvanishing AVHE. In the piezoelectric case \cite{Kalameitsev2019}, in addition to the warping effect of Fermi surface, the nontrival Berry phase can also give rise to AVHE.

In this paper, we report a new contribution to the nonlinear AVHE: the tilting effect of Dirac cones. We will show that the nonlinear AVHE does emerge even in the complete absence of warping electron dispersion and without considering the Berry phase in the 2D titled Dirac system.
\label{Review of VAE}

\emph{Acoustoelectric effect in 2D Dirac material}. The formulas for nonlinear current generated from the SAW via AEE have been recently determined through the semiclassical framework of electron dynamics~\cite{Kalameitsev2019}. We start by recalling the formulas. When propagating a Bleustein-Gulyaev SAW with wave vector $\mathbf{k}$ and frequency $\omega$ along the interface of the 2D materials and the piezoelectric substrate, an in-plane piezoelectric field $\mathbf{E}\left(\mathbf{r},t\right)=\text{Re}\left(\mathbf{E}e^{i\mathbf{k}\cdot\mathbf{r}-i\omega t }\right)$ will be created.
Meanwhile, an induced electric field $\mathbf{E}^i\left(\mathbf{r},t\right)$ stemmed from the fluctuations of the electron density will also emerge owing to the perturbation of SAW, which can be determined through Maxwell's equations (see details in Ref.\cite{SM}). Subsequently, the overall electric field $\widetilde{\mathbf{E}}\left(\mathbf{r},t\right)$, which includes the in-plane piezoelectric field $\mathbf{E}\left(\mathbf{r},t\right)$ and the induced electric field $\mathbf{E}^i\left(\mathbf{r},t\right)$, will drag the carriers in 2D materials, giving rise to a nonlinear current.
The nonlinear current in response to the SAW-induced electric field can be formally decomposed into drift and  diffusive components ${j}_{a}=j_a^\text{dr}+j_{a}^\text{di}$ with $j_a^\text{dr}=\text{Re}\left(\chi^\text{dr}_{abc}\widetilde{E}_b^{*}\widetilde{E}_c\right)$ and $j_a^\text{di}=\mathrm{Re}\left({\chi^\text{di}_{abc}}
\widetilde{E}^{*}_b\widetilde{E}_c\right)$ (the superscripts ``{dr}" and ``{di}" refer to drift and diffusive, respectively).
The response functions $\chi^\text{dr}_{abc}$ and $\chi^\text{di}_{abc}$ have forms
\begin{equation}
\begin{aligned}
\chi^\text{dr}_{abc}&=-2e^3\tau^2Q_{abc},\\
\chi^\text{di}_{abc}&=-2e\tau\frac{\partial \mu}{\partial n}P_{ab}k_d\sigma_{dc}/\left(\omega-\mathbf{k}\cdot\mathbf{R}\right),
\end{aligned}
\label{Chi}
\end{equation}
where $a, b, c, d\in \{x,y\}$,  ${\tau}$ represents the scattering time,
${\mu}$ refers to the chemical potential, $n$ denotes the electron density, $\mathbf{R}$ and $\sigma_{dc}$ indicate diffusion vector and conductivity tensor, respectively, which formulas are given in Ref.\cite{SM}, and the pseudotensorial quantities $Q_{abc}$ and $P_{ab}$ are defined, respectively, as
	\begin{align}
	Q_{abc}=\frac{1}{2\hbar}\int\frac{\mathrm{d}\mathbf{p}}{(2\pi)^2}\frac{\partial v_a}{\partial p_b}\frac{v_c}{1-i(\omega-\mathbf{k}\cdot\mathbf{v})\tau}\left(-\frac{\partial f(\varepsilon_{\mathbf{p}})}{\partial \varepsilon_{\mathbf{p}}}\right) \label{Q-abc} \\
	P_{ab}=\frac{1}{2\hbar}\int\frac{\mathrm{d}\mathbf{p}}{(2\pi)^2}\frac{\partial v_a}{\partial p_b}\frac{1}{1-i(\omega-\mathbf{k}\cdot\mathbf{v})\tau}\left(-\frac{\partial f(\varepsilon_{\mathbf{p}})}{\partial \varepsilon_{\mathbf{p}}}\right),\label{P-ab}
	\end{align}
where $\mathrm{\hbar}$ is the Planck constant, ${\varepsilon_\mathbf{p}}$ is the energy of electron with momentum $\mathbf{p}$, $\mathbf{v}=\partial \varepsilon_\mathbf{p}/\hbar\partial \mathbf{p}$ is the velocity of electron, and $f(\varepsilon_{\mathbf{p}})$ indicates the equilibrium Fermi-Dirac distribution function in the absence of the perturbation of SAW.
\emph{Model}. the effective Hamiltonian of  tilted Dirac systems is
\begin{align}
		H_d=v_{F}\hbar(\eta p_x \sigma_{x}+p_{y}\sigma_{y})+\sigma_{z}\Delta/2+\eta t p_x, \label{Hami}
\end{align}
where $\hat{\boldsymbol{\sigma}}$ denotes the Pauli matrices for the two basis function of energy bands, $\eta=\pm1$ indicates the valley index, $\Delta$ presents the energy gap, and $t$ is the tilting parameter. For simplicity, we only focus on the n-doped system, the energy eigenvalue of conduction band is
\begin{equation}
\varepsilon_{\mathbf{p}}=\sqrt{(\frac{\Delta}{2})^{2}+(v_{F}\hbar p)^2}+\varepsilon_t,
\label{energy}
\end{equation}
where $\varepsilon_t=\eta t p_x$ is the tilt-induced energy shift. The band structures with and without tilting effect are illustrated in Fig.~\ref{Fig1}(b). 
The partial derivative of Fermi-Dirac distribution $f\left(\varepsilon_\mathbf{p}\right)$ function with the respect to energy $\varepsilon_\mathbf{p}$ in Eqs.~\eqref{Q-abc}\eqref{P-ab} to the first order of tilting effect can be written as
\begin{equation}
\frac{\partial f(\varepsilon_{\mathbf{p}})}{\partial \varepsilon_\mathbf{p}}=\frac{\partial f_0}{\partial \varepsilon_\mathbf {p}^0}+\varepsilon_t\frac{\partial^2 f_0}{\partial (\varepsilon_\mathbf{p}^0)^2},
\label{Fermi}
\end{equation}
where  $\varepsilon_\mathbf {p}^0=\sqrt{(\frac{\Delta}{2})^{2}+(v_{F}\hbar p)^2}$ is energy without tilting effect. Combining Eqs.~\eqref{Chi}\eqref{Q-abc}\eqref{P-ab} with Eqs.~\eqref{energy}\eqref{Fermi}, the total nonlinear current can determined and would be decomposed into two parts as $\mathbf{j}^\text{total}=\sum_{\eta}\mathbf{j}_{\eta}^\text{tilt}+\mathbf{j}_\text{c}^\text{di}$ corresponding to the tilt-induced valley dependent current $\mathbf{j}_{\eta}^\text{tilt}$ and conventional diffusive current $\mathbf{j}_\text{c}^\text{di}$ with the subscript ``{c}" (superscript ``{di}") referring
to conventional (diffusive), respectively~\cite{SM}.  The conventional diffusive current $\mathbf{j}_{\mathrm{c}}^\text{di}$ is found to be collinear with SAW and does not depend on the valley index and the tilting effect, having no contribution to valley Hall current~\cite{SM}. Thus, we will ignore this conventional diffusive current when studying the nonlinear acoustic valley Hall effect in the following.

When propagating SAW along $\vec{e}_\mathbf{k}=(\cos\theta,\sin\theta)$ direction, where azimuth angle $\theta$ is measured from the tilting direction ($x$-direction), 
the tilt-induced valley dependent current $\mathbf{j}^\text{tilt}_{\eta}$ for $\eta$ valley in the $x$ and $y$ direction as the response to SAW-induced field are found to be~\cite{SM}, respectively,

\begin{align}
	\begin{pmatrix}
		j^\text{tilt}_{\eta,x} \\
		j^{\text{tilt}}_{\eta,y}
	\end{pmatrix}=\eta \Upsilon^{\text{tilt}}_\text{A}
	\begin{pmatrix}
		2+\mathrm{cos2\theta} \\
		\mathrm{sin2\theta}
	\end{pmatrix}E^{2}_{0},
\label{jvdr(jvdi)}
\end{align}
where $E_{0}=k\varphi_\text{SAW}$ is the piezoelectric field amplitude linearly dependent on the magnitude of the SAW wave vector $k$ and the acoustic wave piezoelectric potential amplitude $\varphi_\text{SAW}$, and the valley independent nonlinear current response function (NCRF) amplitude  $\Upsilon^{\text{tilt}}_\text{A}=\Upsilon^\text{dr}_\text{A}+\Upsilon^\text{di}_\text{A}$ (the subscript ``{A}" represents the amplitude) is the sum of the tilt-induced drift NCRF amplitude $\Upsilon^{\text{dr}}_\text{A}$ and tilt-induced diffusive NCRF amplitude $\Upsilon^\text{di}_\text{A}$, which are given by, respectively,
\begin{equation}
	\begin{aligned}
		\Upsilon^{\text{dr}}_\text{A}&=-\frac{te^3\tau^2}{16\pi\hbar^3}\gamma_0\frac{1+\left(\sigma/\sigma_*\right)^2\left(ak \varepsilon_{1}/E_f\right)^2}
		{1+\left(\sigma/\sigma_*\right)^2\left(1+ak\varepsilon_{1}/E_f\right)^2},\\
		\Upsilon^\text{di}_\text{A}&=\frac{te^3\tau^2}{16\pi\hbar^3}\gamma_0 \frac{\left(\sigma/\sigma_*\right)^2\left(ak \varepsilon_{1}/E_f\right)^2}{1+\left(\sigma/\sigma_{*}\right)^2\left(1+ak\varepsilon_{1}/E_f\right)^2}, \label{jAdrjAdi}
	\end{aligned}
\end{equation}
leading to
\begin{equation}
	\Upsilon^{\text{tilt}}_\text{A}=-\frac{te^3\tau^2}{16\pi\hbar^3}\gamma_0\frac{1}
	{1+\gamma(E_f,\tau)},
	\label{up-total-1}
\end{equation}
where the auxiliary function $\gamma_0=1+2\varsigma^2-3\varsigma^4$ is determined through  $\varsigma=\Delta/2E_f$, $\gamma(E_f,\tau)=\left(\sigma/\sigma_*\right)^2\left(1+ak\varepsilon_{1}/E_f\right)^2$ with  $\varepsilon_{1}={m_{e}v_{F}^{2}/}{2}$,
two defined parameters $\sigma_*=\epsilon_0(\epsilon+1)v_s/4\pi$ and  $a=\epsilon_0(\epsilon+1)\hbar^2/(2m_ee^2)$ are dependent on dielectric permittivity of vacuum
$\epsilon_0$ and dielectric constant $\epsilon$ of substrate with sound velocity $v_s$ \cite{Sonowal2020} and the free electron mass $m_e$, and $\sigma=e^2\tau E_f\left(1-\varsigma^2\right)/2\pi \hbar^2$ denotes the static conductivity of the system.
 The formula of tilt-induced NCRF amplitude $\Upsilon_\mathrm{A}^\mathrm{tilt}$ in Eq.~(\ref{up-total-1}) can be further simplified in the following two limits  of $\gamma$ as
\begin{equation}
	\Upsilon_\mathrm{A}^\mathrm{tilt}\approx
	\begin{cases}
		-\frac{t\pi\hbar}{e\Delta^2}\frac{\varsigma^2(1+3\varsigma^2)}{(1-\varsigma^2)\left(1+ak\varepsilon_{1}/E_f\right)^2}\sigma_{*}^2,& \gamma(E_{f},\tau)\gg1,\\
		-\frac{te^3\tau^2}{16\pi\hbar^3}\gamma_0,& \gamma(E_{f},\tau)\ll1,\label{gammaAtilt-sigma}
	\end{cases}
\end{equation}
showing the following relaxation-time dependence: $\Upsilon^\text{tilt}_\text{A}\propto \tau^{2}$ when $\gamma(E_{f},\tau)\ll 1$ and $\Upsilon^\text{tilt}_\text{A} \propto \tau^{0}$ when $\gamma(E_{f},\tau)\gg 1$. It's should be pointed out that $\gamma(E_{f},\tau)\ll 1$ corresponds to the highly disordered and low-doped systems, whereas $\gamma(E_{f},\tau)\gg 1$ can be valid if one does not consider the highly disordered and low-doped systems~\cite{Kalameitsev2019}.
Based on the relaxation-time independence in the regime where $\gamma(E_{f},\tau)\gg1$, the tilting contribution to the acoustic valley Hall effect can be easily distinguished from the Berry phase or trigonal warping~\cite{Kalameitsev2019}, since the Berry-phase induced AVHC has been found to be inversely proportional to relaxation time $\tau$ and AVHC from trigonal-warping contribution has $A+B\tau^{2}$ dependence on the relaxation time $\tau$. Therefore, one could separate the tilt-induced AVHC from the Berry-phase-induced one through the scaling relation: $j^\text{tilt}\propto \rho_{xx}^{0}$ and $j^\text{BP}\propto \rho_{xx}$, where the superscripts ``{BP}" represents Berry phase and $\rho_{xx}$ denotes the longitudinal resistivity.  

According to Eq.~\eqref{jvdr(jvdi)}, one can observe that although the total tilt-induced nonlinear current $\mathbf{j}^\text{tilt}_{\text{total}}=\mathbf{j}^{\text{tilt}}_{\eta=+1}+\mathbf{j}^{\text{tilt}}_{\eta=-1}$ summed over the valley indices $\eta$ is vanishing due to the time reversal symmetry, the valley current
$\mathbf{j}^{\text{tilt}}_{\text{valley}}=\mathbf{j}^{\text{tilt}}_{\eta=+1}-\mathbf{j}^{\text{tilt}}_{\eta=-1}$
 stemmed from the tilting effect is nonzero. Eq.~\eqref{jvdr(jvdi)} also hints that when the SAW is parallel or antiparallel to the tilting direction (i.e., $\theta=0, \pi,$), namely in $x$-direction, valley current manifests itself as a longitudinal current in response to the SAW and there is no valley Hall current flowing vertically to SAW since only the $x$ component of $\mathbf{j}^{\text{tilt}}_{\text{valley}}$, which is aligned to SAW, is nonzero. When the SAW propagates perpendicularly to the tilting direction
(i.e., $\theta=\pi/2, 3\pi/2$), the valley current still only has a nonzero component in $x$-direction but behaves as a valley transverse current (i.e., valley Hall current) since the current, actually, flows vertically to SAW at this situation. 

Actually, the angular dependence of nonlinear longitudinal  current $j^{\text{tilt}}_{\eta,\parallel}$ collinear with the SAW and transverse current $j^{\text{tilt}}_{\eta,\perp}$ vertically to the SAW for valley $\eta$ as the second-order response to SAW-induced field are found to be , respectively~\cite{SM},
\begin{equation}
\begin{aligned}
j^{\text{tilt}}_{\eta,\parallel}&=\cos\theta j^{\text{tilt}}_{\eta,x}+\sin\theta j^{\text{tilt}}_{\eta,y}=3\eta \cos\theta\Upsilon^{\text{tilt}}_{\text{A}}E^{2}_{0},\\
j^{\text{tilt}}_{\eta,\perp}&=-\sin\theta j^{\text{tilt}}_{\eta,x}+\cos\theta j^{\text{tilt}}_{\eta,y}=-\eta \sin\theta\Upsilon^{\text{tilt}}_{\text{A}}E^{2}_{0}, \label{vhc}
\end{aligned}
\end{equation}
showing that the amplitude of nonlinear longitudinal current $j^{\text{tilt}}_{\eta,\parallel}$ aligned to the SAW is triple of  that of nonlinear transverse current $j^{\text{tilt}}_{\eta,\perp}$ which flows vertically to SAW, namely the nonlinear Hall current. Therefore, the SAW driven nonlinear acoustic valley Hall current (AVHC) $j_\text{H}^{\text{valley}}$ stemmed from tilting effect in the tilted Dirac system is
\begin{equation}
j_\text{H}^{\text{valley}}=j^{\text{tilt}}_{+1,\perp}-j^{\text{tilt}}_{-1,\perp}=
-2\sin\theta \Upsilon^{\text{tilt}}_{\text{A}}  E^{2}_{0}.
\label{J-H}
\end{equation}

\begin{figure}[ht]
	\centering
	\subfigure{
		\includegraphics[width=0.99\linewidth]{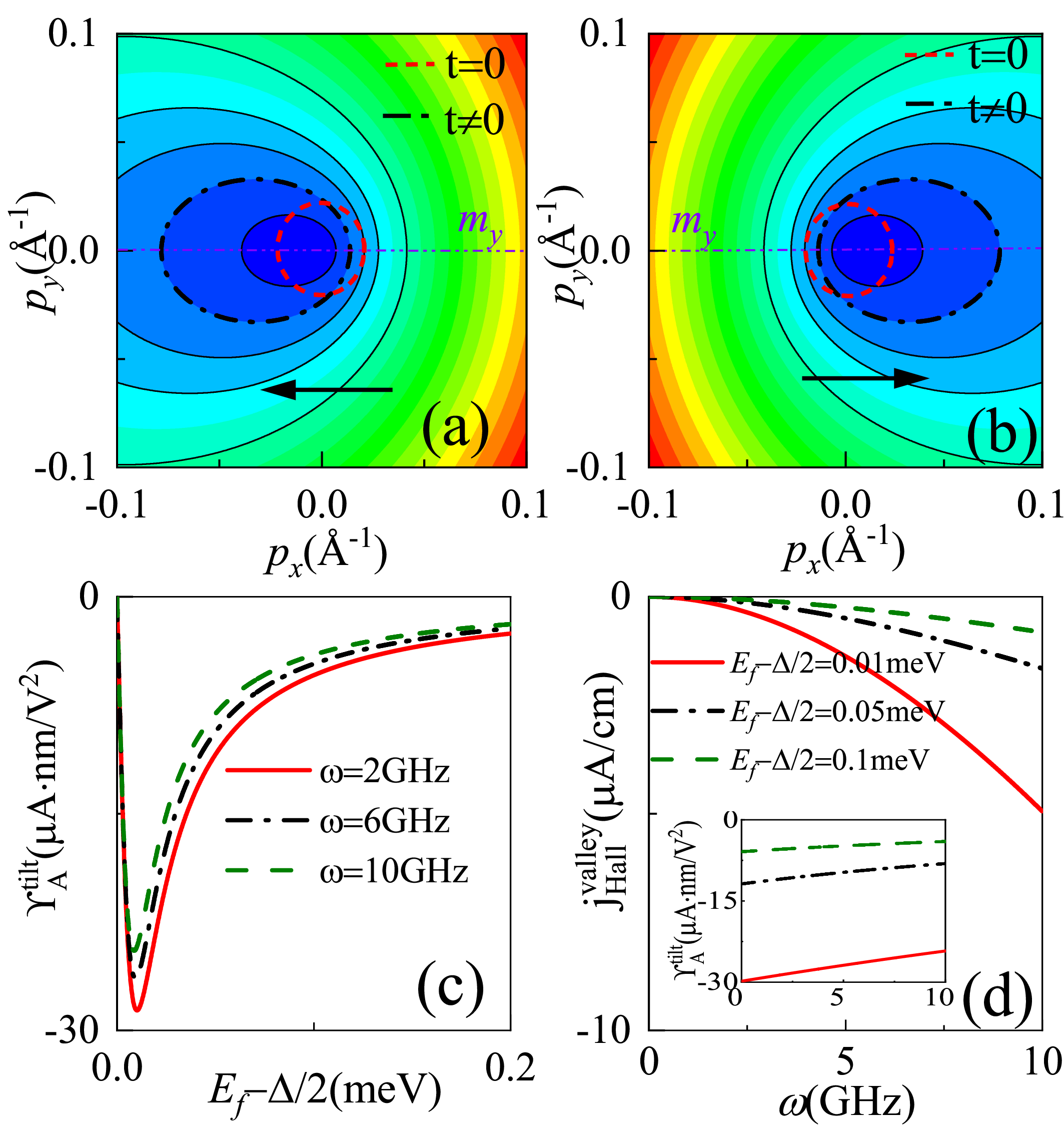}}
	\caption{(a)(b) Schematic of energy contour of K valley [(a)] and -K valley [(b)] with or without tilting effect. The contours show that the mirror symmetry $m_y$ is survived when the tilt is along $x$-direction. (c) $\Upsilon^\text{tilt}_\text{A}$ versus the Fermi energy $E_F-\Delta/2$ at different frequency $\omega$ of SAW. (d) Dependence of the tilt-induced  nonlinear AVHC $j^\text{valley}_\text{Hall}$ on SAW frequency $\omega$. The inset shows the frequency dependence of the $\Upsilon^\text{tilt}_\text{A}$. The black arrows in (a) and (b) represent the tilting direction.
  Parameters used here: $\Delta=5\text{meV}$~\cite{SongPRL2013}, $v_F=0.8\times10^6\text{m/s}$~\cite{fan2011tunable}, and $t=0.03v_F\hbar$ for the uniaxial strain of order $5\%$.}\label{Fig.2}
\end{figure}

Equations \eqref{up-total-1} and \eqref{J-H} show that  $j_\text{H}^{\text{valley}}$ exhibits a $\sin\theta$ dependence on the orientation of SAW with respect to the tilting direction and is proportional the tilting parameter component in the vertical direction to SAW, namely $t\sin\theta$. Thus, when the tilt is perpendicular to the SAW (i.e., $\theta=\pi/2,3\pi/2$), the magnitude of $|j_\text{H}^{\text{valley}}|$ will reach its maximum. However, once the tilt is aligned to the direction of SAW, namely $\theta=0$ or $\pi$, the acoustic nonlinear valley Hall current $j_\text{H}^{\text{valley}}$ vanishes.

The disappearing $j_\text{H}^{\text{valley}}$ can be attributed to the mirror reflection symmetry. Essentially, the tilt does not break the mirror reflection symmetry of each valley along the direction vertical to the tilt [Fig.~\ref{Fig.2}(a)(b)]. The survived mirror symmetry requires  no nonlinear current flowing perpendicularly to the tilt, meaning no acoustic nonlinear Hall current generated. To understand the restriction of mirror symmetry on the nonlinear current orthogonal to the mirror plane, let's assume the tilt is along $l$ direction. Hence, the mirror symmetry $m_{l_{\perp}}$ is survived. Under mirror symmetry $m_{l_\perp}$, $v_{l_{\perp}}$, $p_{l_{\perp}}$ and $\tilde{E}_{l_\perp}$  change sign while $\varepsilon_{\mathbf{p}}$, $p_{l}$, and $\tilde{E}_{l}$ are invariant, where ${l_\perp}$ indicates a vector orthogonal to the vector $l$ in the 2D plane.
Hence, when the system is invariant under mirror symmetry $m_{l_{\perp}}$, the integration of nonlinear acoustic current $j_{l_{\perp}}=-\frac{e^{2}\tau}{\hbar}\text{Re}\int \frac{d\mathbf{p}}{(2\pi)^{2}} v_{l_{\perp}} \tilde{\mathbf{E}}\cdot\frac{\partial f_{1}}{\partial \mathbf{p}}$ in $l_{\perp}$ direction for each valley is an odd function with respect to $p_{\perp}$, hinting that there is no acoustic nonlinear valley current generation vertical to the mirror plane ($j_{l_{\perp}}=0$).

One  candidate Dirac material to observe the predicted nonlinear AVHE stemmed from the tilting effect is the armchair uniaxially strained graphene monolayer. When applying a slight uniaxial strain $u_{yy} (<5\%)$ along the armchair, the tilting parameter $t$ can be determined by $t=0.6u_{yy}v_F\hbar$, and meanwhile the strain-induced anisotropy of the Fermi velocity would be rarely taken into account \cite{RMP2011,PRB2018,das2023NVHE}. Besides, further depositing a hexagonal boron nitride (h-BN) dielectric layer between the piezoelectric substrate and graphene~\cite{Giovannetti2007PRB,huntscience2013,SongPRL2013,zomer2011APL}, a staggered chemical potential $\Delta$ (Semeoff mass, or energy gap) can be generated and the gap values as larger as $\Delta\approx5\sim40~meV$ can be realized\cite{SongPRL2013}. When taking $\Delta=5 meV$, the effective mass of electron $m\approx\Delta/(2v_{F})^{2}$ near the Dirac cone is $7\times10^{-4}m_e$  with $m_e$ presenting the free electron mass ~\cite{SongPRL2013}. The mobility $\mu_{e}$ of the graphene placed on the h-BN layer ranges from $8\times10^4$ to $2.75\times10^5$~$\mathrm{cm^2V^{-1}s^{-1}}$~\cite{zomer2011APL} at low density ($n<10^{-10} cm^{-2}$).  We take $\mu_e$=$2.7\times10^5$~$\mathrm{cm^2V^{-1}s^{-1}}$. Therefore, the scattering relaxation time $\tau=0.1\text{ps}$ is estimated by $\tau=\mu m/e$.  To numerically analyze the behaviours of the tilt-induced AVHE in the uniaxial strained graphene, we choose LiNbO$_{3}$ as the piezoelectric substrate and the corresponding material parameters are taken as follows: the sound velocity $v_s=3500~m/s$, the dielectric constant $\epsilon=50$~\cite{Savenko2020}, and  the acoustic wave piezoelectric potential amplitude $\varphi_{\mathrm{SAW}}=50~\mathrm{mV}$ which determines the amplitude of piezoelectric field $E_{0}=k\varphi_{\mathrm{SAW}}$ ($k=\omega/v_{s}$).

Figure~\ref{Fig.2}(c) shows the dependence of the tilt-induced NCRF amplitude $\Upsilon^\text{tilt}_\text{A}$ on the Fermi energy $E_{f}$ with different frequencies. Obviously, the maxima of tilt-induced NCRF amplitude $\Upsilon^\text{tilt}_\text{A}$ can be obtained by modulating the Fermi energy close to the Dirac point within $0.02~\mathrm{meV}$ through the gate voltage. Therefore, the tilt-induced nonlinear AVHC can be easily separated from the warping effect since the warping-effect contribution becomes significant to nonlinear AVHC only when the Fermi energy is far away from the Dirac point. Although the magnitude of peak value of $\Upsilon^\text{tilt}_\text{A}$ enhance with decreasing the frequency [Figs.~\ref{Fig.2}(c)(d)], the tilt-induced AVHC $\mathrm{j_{H}^{valley}}$ increase when enhancing the frequency owing to  $\mathrm{j_{H}^{valley}\sim}\Upsilon^\text{tilt}_\text{A}\omega^2$ [Figs.~\ref{Fig.2}(d)].
To estimate the tilt-induced acoustic nonlinear valley Hall effect, we take $\Upsilon^\text{tilt}_\text{A}=\mathrm{24.2\mu A\cdot nm /V^2}$ at $E_f-\Delta/2=0.01~\mathrm{meV}$ and $\omega=10~\text{GHz}$. Hence, the pure AVHC $j^{\text{valley}}_{H}=\Upsilon^\text{tilt}_\text{A}E^{2}_{0}$ is estimated to be $4.9\times10^3~\mathrm{nA /cm}$, which is two orders of magnitude greater than that from the Berry phase effect and the warping effect~\cite{Kalameitsev2019}. To detect the predicted pure AVHC here, one would apply the nonlocal resistance measurement in experiment, which has been widely used for the system without valley polarization ~\cite{gorbachev2014detecting,sui2015gate,shimazaki2015generation,wu2019intrinsic}.

\label{Conclusions}
\emph{Conclusions}.
We show that a nonlinear acoustic valley Hall effect emerges  in tilted Dirac systems in complete absence of warping effect and without considering Berry phase. It's found that the  nonlinear acoustic valley Hall effect has a contribution from tilting effect and show a $\sin\theta$ dependence on the orientation of tilt with respect to the  surface acoustic wave. Interestingly , the tilt-induced nonlinear acoustic valley Hall effect shows a relaxation-time independence in the regime $\gamma(E_{f},\tau)\gg 1$, which presents an approach to distinguish the contributions from the Berry phase or trigonal warping. We have also calculated the nonlinear acoustic valley Hall effect in the armchair uniaxially strained graphene monolayer with substrate-induced energy gap. Remarkably, the magnitude of nonlinear AVHC stemmed from tilt mechanism in the strained graphene is two orders larger than those reported arisen from warping effect and Berry phase in monolayer transition metal dichalcogenides.

\emph{Acknowledgements}.
This work is supported by  the NSFC  (Grant No.12004107 and No.12374040), the National science Foundation of Hunan, China (Grant No.2023JJ30118), and the Fundamental Research Funds for the Central Universities.

\end{document}